\documentclass[12pt,preprint]{aastex}
\input epsf.tex

\newcommand{\alphab}{\mbox{\boldmath$\hat{\alpha \;} $}}
\newcommand{\deltab}{\mbox{\boldmath$\hat{\delta \;} $}}

\begin{document}
 
\title{THE GALACTIC INNER HALO: SEARCHING FOR WHITE DWARFS AND MEASURING THE 
FUNDAMENTAL GALACTIC CONSTANT, $\Theta_\circ$/$R_\circ$$^{\rm 1}$}

\author{
Jasonjot Singh Kalirai\altaffilmark{2},
Harvey B. Richer\altaffilmark{2},
Brad M. Hansen\altaffilmark{3},
Peter B. Stetson\altaffilmark{4},   
Michael M. Shara\altaffilmark{5},
Ivo Saviane\altaffilmark{6},
R. Michael Rich\altaffilmark{3},
Marco Limongi\altaffilmark{7},
Rodrigo Ibata\altaffilmark{8},
Brad K. Gibson\altaffilmark{9},
Gregory G. Fahlman\altaffilmark{4} \&
James Brewer\altaffilmark{2}}

\shorttitle{WHITE DWARFS IN THE GALACTIC INNER HALO}
\shortauthors{Kalirai {\it et al.}}

\altaffiltext{1} 
{Based on observations with the NASA/ESA Hubble Space Telescope, obtained at
the Space Telescope Science Institute, which is operated by AURA under NASA
contract NAS 5-26555. These observations are associated with proposal GO-8679.}
\altaffiltext{2} 
{Department of Physics \& Astronomy, University of British Columbia,
6224 Agricultural Road, Vancouver, BC V6T 1Z1, Canada. jkalirai@astro.ubc.ca, 
richer@astro.ubc.ca, jbrewer@astro.ubc.ca}
\altaffiltext{3} 
{Department of Physics \& Astronomy, University of California at Los Angeles,
Math-Sciences 8979, Los Angeles, CA 90095-1562. hansen@astro.ucla.edu, 
rmr@astro.ucla.edu}
\altaffiltext{4} 
{National Research Council of Canada, Herzberg Institute of Astrophysics, 5071 West
Saanich Road, RR5, Victoria, BC V9E 2E7, Canada. peter.stetson@nrc.gc.ca, 
Greg.Fahlman@nrc.gc.ca}
\altaffiltext{5} 
{American Museum of Natural History, Astrophysics Department, Central Park 
West \& 79th Street, New York, NY 10024-5192. mshara@amnh.org}
\altaffiltext{6} 
{European Southern Observatory, Alonso de Cordova 3107, Vitacura, Casilla 
19001, Santiago 19, Chile. isaviane@eso.org}
\altaffiltext{7} 
{Osservatorio Astronomico di Roma, Via Frascati 33, I-00040 Monte Porzio 
Catone, Rome, Italy. marco@nemo.mporzio.astro.it}
\altaffiltext{8} 
{Observatoire de Strasbourg, 11 rue de l'Universite, F-67000 Strasbourg, 
France. ibata@newb6.u-strasbg.fr}
\altaffiltext{9} 
{Centre for Astrophysics and Supercomputing, Swinburne University, Mail 31, 
P.O. Box 218, Hawthorn, Victoria, 3122 Australia. bgibson@astro.swin.edu.au}

\vspace{.2in}

\begin{abstract} 
We establish an extragalactic, zero-motion frame of reference within the 
deepest optical image of a globular star cluster, a {\sl Hubble Space 
Telescope (HST)} \rm 123-orbit exposure of M4 (GO 8679, cycle 9).  The 
line of sight beyond M4 ($l$, $b$ = 351$^{\rm o}$, 16$^{\rm o}$) intersects 
the inner halo (spheroid) of our Galaxy at a tangent-point distance of 7.6 kpc 
(for $R_\circ$ = 8 kpc).  The main sequence of this population can be clearly 
seen on the color-magnitude diagram (CMD) below the M4 main sequence.  
We isolate these spheroid stars from the cluster on the basis of their proper motions
over the 6-year baseline between these observations and others made at a 
previous epoch with {\sl HST} \rm (GO 5461, cycle 4).  Distant background 
galaxies are also found on the same sight line by using image-morphology 
techniques.  This fixed reference frame allows us to determine an independent 
measurement of the fundamental Galactic constant, $\Omega_\circ$ = 
$\Theta_\circ/R_\circ$ = 25.3 $\pm$ 2.6 km/s/kpc, thus 
providing a velocity of the Local Standard of Rest $v_{\rm LSR}$ = 
$\Theta_\circ$ = 202.7 $\pm$ 24.7 km/s for $R_\circ$ = 8.0 $\pm$ 0.5 kpc. 
Secondly, the galaxies allow a direct measurement of M4's absolute proper motion, 
$\mu_{\alpha}$\alphab = $\rm-$12.26 $\pm$ 0.54 mas/yr, $\mu_{\delta}$\deltab 
= $\rm-$18.95 $\pm$ 0.54 mas/yr, in excellent agreement with recent studies.  The clear 
separation of galaxies from stars in these deep data also 
allow us to search for inner-halo white dwarfs.  We model the conventional Galactic 
contributions of white dwarfs along our line of sight and predict 7.9 (thin disk), 6.3 
(thick disk) and 2.2 (spheroid) objects to the limiting magnitude at which we can clearly 
delineate stars from galaxies ($V \sim$ 29).  An additional 2.5 objects are expected 
from a 20\% white dwarf dark halo consisting of 0.5 M$_\odot$ objects, 70\% of which 
are of the DA type.  After considering the kinematics and morphology of the objects in our 
data set, we find the number of white dwarfs to be consistent with the predictions 
for each of the conventional populations.  However, we do not find any evidence for 
dark halo white dwarfs.

\end{abstract}

\keywords{Dark matter -- Galaxy: halo, kinematics and dynamics -- galaxies: 
photometry -- globular clusters: individual (Messier 4) -- stars: white dwarfs}

\section{Introduction} \label{intro}

A recent HST imaging project of the Galactic globular cluster M4 (GO 8679) has 
so far investigated both the low-mass end of the hydrogen burning main sequence 
\cite{richer} and the termination of the white dwarf cooling sequence 
\cite{hansen}.  The data for the cluster members was isolated from 
background/foreground contaminants by using the proper motion of the cluster 
with respect to the field over a 6 year baseline (previous epoch data 
obtained with HST in cycle 4, 1995 \cite{richer3,richer2,ibata}).  Richer et al.\ (2002) 
found the M4 cluster mass function to be slowly rising to low masses.  In Hansen et 
al.\ (2002), the cluster white dwarfs were used as chronometers to provide an age 
estimate of 12.70 $\pm$ 0.35 Gyrs (1-$\sigma$) 
largely independent of the turn-off for M4.  In this contribution, we shift our 
focus to the background contamination itself (the spheroid and galaxies) and isolate 
these populations for investigation.  A separate contribution will address the 
main-sequence stars, luminosity and mass functions of the spheroid, and 
foreground thin/thick disk components.

Stars, gas and dust in the disk, bulge and spheroid of our Galaxy 
only account for $\sim$10\% of the total mass within $R$ = 50 kpc 
\cite{wilkinson}.  The remaining mass is believed to reside 
in the dark halo of our Galaxy (dark matter), and determining the 
nature of this mass is a critical issue today in astrophysics.  
Recently, the MACHO project has analyzed data from microlensing events in 
the direction of the LMC and determined that the mean mass of the lenses 
is 0.5 $\pm$ 0.3 $M_\odot$, with a $\sim$20\% MACHO fraction \cite{alcock}.  
This naturally suggests faint white dwarfs as the source of these lenses.  Even 
if the MACHOs cannot account for the total dark matter contribution, determining 
the properties and number density of faint halo white dwarfs is important in 
several areas of astrophysics.  These include studying the IMF of population 
III stars (through their remnants) and constraining the star-formation history 
of our Galaxy.

Studying white dwarfs in the Galactic bulge or spheroid from the ground is difficult 
for two reasons.  First, the end of the white dwarf cooling sequence for 0.5 
$M_\odot$ objects in a population 12 Gyrs old is $M_{V} \sim$ 17 
\cite{hansen2}, which, at the center of the Galaxy (8 kpc), is $V >$ 31.5.  
The depth of the M4 study ($V \sim$ 30), however, is faint enough 
to detect the brighter end of this cooling sequence.  Secondly, piercing 
through the disk of our Galaxy picks up many foreground metal-rich disk 
stars which contaminate the sample.  Although observing at higher latitudes above the 
center helps avoid the thick disk, the resulting spheroid number density also drops 
off rapidly, $n(r) \propto$ $r^{\rm -3.5}$ \cite{binney2}.  With HST we can achieve 
the depth required to measure these faint stars, but there are several disadvantages.  
First, we are dealing with small-number statistics due to the limited field of 
view (for instance, WFPC2 has a field area of 5.7 arcmin$^{\rm 2}$).  More importantly, 
long exposures pick up many 
galaxies, such as seen in the HDF, which can mimic stars in faint photometry.  

Given the caveats listed above, we attempt in this paper to isolate potential 
white dwarfs in the field by separating out populations using both their 
morphology and their kinematics.  After briefly presenting the data in \S 
\ref{data}, we begin by finding galaxies and establishing our 
extragalactic zero-motion frame of reference.  This then allows us to determine 
two important quantitities directly measured from a sample of extragalactic 
objects: the velocity of the Local Standard of Rest (\S \ref{circ}) and the absolute 
proper motion of M4 (\S \ref{absolute}).  
In \S \ref{cmd} we examine the different populations which make up the corrected 
proper-motion and color-magnitude diagrams and discuss the spheroid population.  Next, 
we analyze faint, blue stars in this population and identify our best candidates for disk 
and spheroid white dwarfs (\S \ref{analysis}).  This includes comparing our results to the 
expected numbers of white dwarfs given the searchable volume in our data and the various 
population-density distributions along the line of sight.  In \S \ref{discussion}, we discuss 
the current status of the search for dark halo white dwarfs as well as some of the different 
views which have been presented in the recent literature.  The current results are placed 
in the context of these independent efforts.  The overall study is concluded 
in \S \ref{conclusions}.

\section{The Data} \label{data}

The data (35 hours in F606W, 55 hours in F814W) for the present and previous 
observations (1995 - cycle 4 
\cite{richer3,richer2,ibata}) were reduced using the DAOPHOT/ALLSTAR \cite{stetson1} 
and ALLFRAME \cite{stetson2} photometry packages.  Individual frames were 
registered and co-added using the DAOMASTER and MONTAGE2 programs.  For this we 
transformed all images for a given CCD to a common coordinate system for each 
epoch and filter and combined the images using standard IRAF tasks.  As the fields 
are dominated by M4 cluster members, smaller matching radii rejected non cluster 
members.  A plot showing $dx$ vs $dy$ between the older and newer epochs therefore 
shows the cluster stars grouped at the origin and a secondary group of stars offset 
from this.  The transformations were then iterated using a list of stars likely to 
be cluster members as selected from this clump near the origin.  The CMD of the 
resulting stars verified their cluster origin.  The list of these stars were then 
transformed back onto each of the individual frames which were reduced using ALLSTAR, 
with the recentering option enabled.  The output ALLSTAR files, which contain only 
cluster members, were matched with DAOMASTER using a 20-term transformation equation.  
The 20-term transformation was found to reduce systematic errors significantly from the 
6-term transformation.  We produced histograms of $dx$ and $dy$ for the cluster stars 
and minimized streaming motions near the edges of the CCD caused by distortions by 
reducing the matching radius until it was similar in magnitude to the rms error in 
the transformation.  The residual systematic bias in the proper motion diagram is 
evaluated below.  Finally, MONTAGE2 was used to expand the frames by a factor of 
3 before registering them as noted above.  This procedure is similar to a drizzling 
routine and effectively decreased the cluster proper-motion dispersion in the 
proper-motion diagram.  

The final photometry list includes only those stars which were measured in 
both filters and at both epochs, and which passed a visual inspection.  Further 
details of the reduction and calibration of the data set used in this analysis 
will be described in Richer et al.\ (2003) (see also \S 2 of Richer et al.\ (2002)).  We 
summarize the key parameters for both the cluster and field population and present 
the observational log in Table 1.

In Figure 1, we present the proper-motion diagram for our data.  The motion of the 
cluster with respect to the field (6 year baseline), can be clearly seen as the tight 
clump toward negative proper motions.  For the units, we first translated the $x$ and $y$ 
pixel motions into Galactic coordinates ($l$, $b$) by using a rotation angle which aligned 
the $y$ axis of the CCD to the North Galactic Pole.  To express these proper motions, we 
adopt the convention that the vector proper motion is given by $\mu$ = 
$\sqrt{(\mu_{l}{\bf \hat{l}})^{2} + (\mu_{b}{\bf \hat{b}})^{2}}$, where ${\bf \hat{l}}$ and 
${\bf \hat{b}}$ are unit vectors in the corresponding directions.  We then converted the HST 
WFPC2 pixels into arcseconds using the plate scale (0\farcs1/pixel, Biretta et al.\ 2002) of 
the CCD.  The motions are then divided by the baseline of 6 years and converted to mas/yr.  
Zero motion in the diagram, ($\mu_l$${\bf \hat{l}}$, $\mu_b$${\bf \hat{b}}$) = (0, 0), is 
centered on well measured, bright galaxies (larger red dots) and is described in detail 
in \S \ref{centering}.  A small systematic bias, seen as a stretching of the proper-motion 
diagram in the $l$ direction, is evaluated to be 0.26 mas/yr.  This was measured by producing 
histograms of the proper-motion diagram for different magnitude cuts.  Comparing the 
dispersions in the $l$ and $b$ directions for the brightest stars directly gives the error.  
This systematic bias has been factored into the error budget for all measurements.

\section{An Extragalactic Reference Frame} \label{galaxies}

\subsection{Measuring Galaxies} \label{expectedgalaxies}

Isolating galaxies in the current study is crucial as they can mimic faint blue 
white dwarfs in the data.  Additionally, since the galaxies are not moving, they 
represent a fixed zero-motion position in the proper-motion diagram from which we can measure 
absolute motions.  Visually, the images of M4 only show a small number of obvious 
galaxies.  Were it not for the greatly larger number of foreground stellar 
objects, the 1.3 magnitudes of foreground visual extinction and the higher 
background produced by scattered light and zodiacal light, this image would likely 
look similar to the HDF \cite{williams} which shows 1781 galaxies in the range 
26 $\leq V_{\rm AB}$ $\leq$ 29.5. 

To measure galaxies in our field, we used the image morphology classification tool, 
SExtractor \cite{bertin}.  SExtractor assigns a stellarity 
index to each object which can be used to distinguish between stars and galaxies.  
The stellarity is determined through a neural network, which learns based on 
other, high signal-to-noise, stars. Although SExtractor was not able to recover 
statistics for every object that ALLSTAR measured in our data, the classifications 
are $\sim$75\% complete through $V$ = 24--29.  Fainter than $V$ = 29, SExtractor was 
only able to measure $\sim$40\% of the ALLSTAR sources and struggled to classify 
objects as stars or galaxies.  Altogether, just over 50 galaxies were identified and 
measured using a stellarity $<$ 0.2 cut.

Figure 2 (top) shows the SExtractor classifications as a function of magnitude for all 
objects.  The distribution of points on this diagram can 
be broken into two classes which are separated by the horizontal lines.  The stars are found 
predominantly in a clump near stellarity = 1 down to a magnitude 
of $V \sim$ 28.  Fainter than this, the classifications are not as good.  However a 
clump of stars can be seen extending to $V \sim$ 29 with a sharp cutoff in stellarity 
at 0.7.  The galaxies are found near the bottom of the figure, at stellarity = 0.  
Here we have chosen a conservative stellarity $<$ 0.2 cut to include possible outliers 
above the predominant galaxy sequence.  Very few objects ($\sim$4\% for V $<$ 29) are seen 
between the stars and galaxies indicating that the classifications are reliable.  At the 
faintest magnitudes, $V \sim$ 29, a clump of objects can be seen at stellarity = 0.5.  This is 
expected as the program will, on average, choose stellarity = 0.5 for an object which 
it cannot classify.  Shape parameter comparisons from artificial star tests also confirm that 
the stellarity is measured accurately down to this limit.  Therefore $V \sim$ 29 represents 
the limit at which we can accurately separate stars from galaxies.

In Figure 2 (bottom) we use the stellarity to illustrate the separate populations 
in our data.  Here, the stellarity of all objects is plotted against the total absolute 
proper motion, $\mu_{\rm total}$ = $\sqrt{(\mu_{l}{\bf \hat{l}})^{2} + (\mu_{b}{\bf \hat{b}})^{2}}$ 
as determined with respect to the center of the large red dots in Figure 1.  We will 
justify this location in the next section.  First, we note two clear clumps for stars 
(at constant stellarity $\sim$ 1) representing the cluster ($\mu_{\rm total} \sim$ 22.5 mas/yr, 
with a small dispersion) and field populations ($\mu_{\rm total} \sim$ 5 mas/yr, with a 
much larger dispersion).  This bi-modality continues for lower stellarities down to 
stellarity $\sim$ 0.7.  The galaxies (stellarity $\sim$ 0) are obviously found at low 
$\mu_{\rm total}$.  However, a tail is seen to higher proper-motion displacements (these 
are the small red dots in Figure 1).  As we will see in \S \ref{centering}, this tail 
represents the faintest galaxies and indicates that our ability to measure the proper 
motions has degraded due to reduced signal-to-noise.

\subsection{Centering the Zero-Motion Frame of Reference} \label{centering}

In Figure 1, we showed the locations of bright, well-measured galaxies (large red 
dots) which represent an absolute background reference frame.  As these galaxies 
clumped together, we first estimated their position and measure the absolute proper 
motion of all objects with respect to this point.  

Figure 3 shows the total proper-motion displacement as a function of magnitude for both 
stars (top) and galaxies (bottom).  The distributions are found to be very different.  The 
stars behave as expected and are confined within a constant envelope in $\mu_{\rm total}$ 
across all but the faintest magnitudes.  Note that the spread in the distribution of M4 
cluster members ($\mu_{\rm total} \sim$ 22.5 mas/yr) for $V \geq$ 28 is not due to the 
main-sequence stars. At $V \gtrsim$ 28, the main sequence of M4 contains very few stars 
\cite{richer} and so the scatter is due to the faintest white dwarfs in the cluster.  
On the other hand, the total proper-motion displacement of the galaxies (bottom) is found 
to completely degrade for $V \gtrsim$ 27.  Considering that we know from Figure 2 that the 
stellarities for the galaxies are measured accurately to well beyond this limit, this flaring 
must be entirely due to difficulties in the astrometric centering on these faint extended 
sources.  Therefore the galaxies with $V \lesssim$ 27 should dictate the location of the 
zero-motion frame as some of their fainter counterparts have been poorly measured.  
The centroid of the 12 galaxies which satisfy this criteria represents the zero-motion 
frame of reference shown on all proper-motion diagrams (see large red dots in Figure 1).  
The one-dimensional error in this location is calculated as $\sigma$/$\sqrt{N} \approx
0.5$~mas/yr, where $\sigma$ is the proper-motion dispersion of the galaxy sample 
($\sim 1.6$~mas/yr).

\subsection{The Circular-Speed Curve} \label{circ}

We now use the extragalactic stationary frame of reference established above to 
measure two important quantities.  The first, the circular-speed curve, is a plot of the 
velocity of a test particle ($\Theta_\circ$) moving in a circular orbit in the Galactic plane and 
around the Galactic center, vs the distance $R$ at which it is located relative to the center.  
The ratio of these quantities at the Solar radius ($\Omega_\circ$ = 
$\Theta_\circ/R_\circ$) represents one of the more difficult problems in Galactic structure 
and directly provides the mass interior to $R_\circ$.  This constant is fundamentally 
important in understanding the dynamics of the Galactic halo and the Local Group.

For the motion in the Galactic plane, we measure an angular proper motion from 
the center of the galaxy clump to the center of the spheroid clump of 
$\mu_l$${\bf \hat{l}}$ = $\rm-$5.53 $\pm$ 0.56 mas/yr.  The spheroid clump was 
chosen by first eliminating foreground disk stars by overplotting a solar metallicity 
disk dwarf fiducial on our CMD (presented in \S \ref{cmd}).  These disk stars represent 
the redder, brighter stars located above M4.  We then removed both M4 and galaxies, and 
iteratively determined the center of the inner halo based on $\sigma$-clipping to remove 
outliers.  The resulting population is shown within the large circle in Figure 1.  
Assuming the spheroid is not rotating (see below), the vector motion of $\mu_l$${\bf \hat{l}}$ 
= $\rm-$5.53 $\pm$ 0.56 mas/yr represents the reflex motion of the Sun, a combination of the 
LSR circular orbit and the deviation of the Sun from that circular orbit (Solar motion).  
The uncertainty in this number is derived from the quadrature sum of the dispersions in each 
of the galaxy and spheroid clumps.  We can correct this observed proper motion for the 
known Solar motion in the Galactic plane, $v_\odot$ = 5.25 $\pm$ 
0.62 km/s in the direction of Galactic rotation \cite{dehnen}.  Orienting this Solar 
motion onto the $l$ direction, we find a correction of $\Delta$$\mu_l$${\bf \hat{l}}$ 
= 0.19 $\pm$ 0.02 mas/yr.  Therefore, $\Omega_\circ$ = $\Theta_\circ$/$R_\circ$ = 
${\rm-}$4.74$\mu_l$${\bf \hat{l}}$ = 25.3 $\pm$ 2.6 km/s/kpc. 

The angular velocity of the circular rotation of the Sun can be directly 
compared to the Oort Constants, $A$ and $B$ \cite{kerr}, which measure the 
shear and vorticity of the disk.  Recent analysis based on {\sl 
Hipparcos} \rm measurements of 220 Galactic Cepheids gives $\Omega_\circ$ 
= $A {\rm-} B$ = 27.19 $\pm$ 0.87 km/s/kpc \cite{feast}, larger than our value 
but consistent within the uncertainty.  Other measurements, such as 
the Sgr A$^{*}$ proper-motion study \cite{reid1}, also find a higher value 
than ours, $\Omega_\circ$ = 27.2 $\pm$ 1.7 km/s/kpc.  At $R_\circ$ = 8.0 
$\pm$ 0.5 kpc, these two measurements give a Solar reflex velocity of 
$v_{\rm LSR}$ = $\Theta_\circ$ $\sim$ 218 km/s, in good agreement with the 
independent IAU adopted value of 220 $\pm$ 15 km/s \cite{binney} based on 
kinematics of high velocity stars.  The most recent study by Bedin et al.\ (2003) 
uses a single QSO in a different field of M4 than studied here and finds 
$\Omega_\circ$ = 27.6 $\pm$ 1.7 km/s/kpc.  However, other studies, such as Kuijken \& 
Tremaine (1994) find much smaller values.  Based on a set of self-consistent 
solutions for different Galactic parameters, they find $v_{\rm LSR}$ = 180 km/s.  
Merrifield (1992) also calculates a lower Local Standard of Rest velocity, $v_{\rm LSR}$ 
= 200 $\pm$ 10 km/s, using rotation of H I layers.  For $R_\circ$ = 8.0 $\pm$ 0.5 kpc, 
our result gives a similar value to this latter study, $v_{\rm LSR}$ = 202.7 $\pm$ 
24.7 km/s.  Only 15\% of the uncertainty in our result is contributed by the 
0.5 kpc uncertainty in the Galactocentric distance with the remaining error almost 
entirely due to the dispersion in the galaxy sample.  The dispersion in the spheroid 
sample is calculated from those stars within the circle in Figure 1 and has also been 
folded into this uncertainty.

Although our value for $\Omega_\circ$ is lower than some previous studies of the circular 
speed curve, it is consistent with those studies within the 1-$\sigma$ uncertainty.  One 
consideration that we are currently investigating is possible rotation by the spheroid 
itself, which would cause an over/underestimate of the Solar reflex motion depending on the 
degree to which stars in front of the tangent point dominate 
over stars behind it.  Given the small field of view, we could also potentially be affected 
by debris tails or small groups of stars with peculiar orbits.   Despite these caveats, 
we stress the importance of this measurement.  Kuijken \& Tremaine (1994) estimate 
that if they are correct and a reduction of $\Theta_\circ$ from 220 km/s to 180 km/s 
is needed, the line-of-sight velocity of M31 relative to the Galactic center would 
correspondingly increase by 25\%.  This, in turn, would increase the mass of the 
Local Group from timing arguments by 35\%.  Although additional methods 
(see Kuijken \& Tremaine 1994 for a summary) to constrain the circular-speed 
curve are being used (e.g., using Local Group kinematics, motions of globular 
clusters or halo stars, tangent-point measurements of the inner rotation curve 
and outer-rotation curve-measurements using OB stars and H II regions, etc.), they are 
not superior to measurements made with extragalactic reference frames.  These global 
measurements, such as the present work and the Sgr A$^{*}$ proper motions \cite{backer,reid1}, 
avoid Galactic variations in the sample of objects and therefore reduce systematic 
uncertainties \cite{binney}. 

The apparent motion of the spheroid stars in the direction perpendicular to the plane 
is very small, $\mu_b$${\bf \hat{b}}$ = $\rm-$0.32 $\pm$ 0.56 mas/yr (see Figure 1).  The 
correction for the component of the known Solar motion, $v_{Z_{\odot}}$ = 7.17 $\pm$ 0.38 km/s 
\cite{dehnen} in this direction is $\Delta$$\mu_b$${\bf \hat{b}}$ = $\rm-$0.12 $\pm$ 0.01 
mas/yr.  Therefore, the residual velocity for $R_\circ$ = 8.0 $\pm$ 0.5 kpc is 
found to be 7.4 $\pm$ 21.2 km/s, consistent with a stationary population.

\subsection{The Absolute Proper Motion of M4} \label{absolute}

The space motions of Galactic globular clusters are of interest in order to constrain 
and reconstruct their orbits around the Galaxy.  This yields important information 
on cluster origins and destruction processes, as well as Galactic dynamics.  Formation 
scenarios of the Galaxy can be better understood by coupling the kinematics of the 
clusters with other properties, such as the metallicities and ages \cite{dinescu1}.

Three primary methods exist to determine the space motions of globular clusters 
and all involve measuring the absolute proper motion of the stars in the cluster.  
The indirect methods involve either using secular parallaxes (which require a 
knowledge of Galactic rotation) or using the new `bulge-relative' method 
(which requires the distance and proper motion of the bulge:  Terndrup et al.\ 
1998).  The preferred method involves using extragalactic objects, galaxies or 
QSOs, and directly measuring the motion of the cluster.  

Only three previous estimates of the M4 absolute proper motion exist in the literature.  
Cudworth \& Rees (1990) used bright field stars as a reference to the brighter cluster stars 
to estimate $\mu_{\alpha}$\alphab = $\rm-$11.6 $\pm$ 0.7 mas/yr, $\mu_{\delta}$\deltab = 
$\rm-$16.3 $\pm$ 0.9 mas/yr.  Dinescu et al.\ (1999) used {\sl Hipparcos} \rm 
field stars as a reference to carefully measure the motion of the cluster after 
accounting for plate-transformation systematics: $\mu_{\alpha}$\alphab 
= $\rm-$12.50 $\pm$ 0.36 mas/yr, $\mu_{\delta}$\deltab = $\rm-$19.93 $\pm$ 0.49 mas/yr.  
The Bedin et al.\ (2003) study finds $\mu_{\alpha}$\alphab = $\rm-$13.21 $\pm$ 
0.35 mas/yr, $\mu_{\delta}$\deltab = $\rm-$19.28 $\pm$ 0.35 mas/yr.  Here we present 
an absolute proper-motion measurement of the cluster measured with respect to 
a sample of galaxies, an important measurement considering that M4 is a very well studied 
cluster.

As discussed earlier, our coordinate differences were measured by centering on the 
cluster stars.  We then found the center of the extragalactic distribution 
and directly measured the difference between this and the 
center of the M4 clump.  Converting Galactic coordinates to the equatorial 
system, we find $\mu_{\alpha}$\alphab = $\rm-$12.26 mas/yr, 
$\mu_{\delta}$\deltab = $\rm-$18.95 mas/yr, with a one-dimensional uncertainty of 
0.54 mas/yr.  The error in the mean for the 
absolute proper motion is the quadrature sum of the total errors for each of the 
galaxy and cluster star distributions.  We've also factored in the 0.26 mas/yr 
systematic bias discussed in \S \ref{data}.  The final errors in 
our absolute proper motion are completely dominated by the dispersion in 
the galaxy sample.  Comparing these results directly to previous studies, we 
find that our total proper motion ($\mu_{\rm total}$ = $\sqrt{(\mu_{\alpha}\alphab)^{\rm 2} 
+ (\mu_{\delta}\deltab)^{\rm 2}}$) is $\sim$4\% smaller than the 
Dinescu et al.\ (1999) value, $\sim$13\% larger than the Cudworth \& Rees (1990) value 
and $\sim$3\% smaller than the Bedin et al.\ (2003) study.  

For completeness, we can also convert our cluster proper motions into absolute space 
motions.  The Solar motion is ($U_\odot$, $V_\odot$, $W_\odot$) = (+10.0, +5.2, +7.2) 
km/s \cite{dehnen}, where $U_\odot$ is positive towards the Galactic center, 
$V_\odot$ is in the direction of Galactic rotation, and $W_\odot$ is out of the 
plane.  Therefore, for M4 we derive an absolute space motion of ($U$, $V$, $W$)$_\circ$ 
= (57.0 $\pm$ 6.3, ${\rm-}$186.1 $\pm$ 6.3, ${\rm-}$2.4 $\pm$ 4.5) km/s with respect to 
the LSR, for a distance of $d_{\rm M4}$ = 1720 pc and a cluster radial velocity of 
$v_{r}$ = 70.9 $\pm$ 0.6 km/s (Peterson, Rees \& Cudworth 1995).  The uncertainties 
calculated for the space motions above do not include any distance uncertainty.

\section{The Proper-Motion and Color-Magnitude Diagrams} \label{cmd}

In Figure 4 we again present the final proper-motion diagram and also the corresponding 
color-magnitude diagram for all objects within the M4 field.  As we mentioned earlier, 
the tighter clump of dots in the proper-motion diagrams 
represent M4 (proper-motion dispersion, $\sigma_{\rm total}$ = 1.9 mas/yr) 
whereas the more diffuse clump represents the spheroid population plus other stars along the 
line of sight (proper-motion dispersion $\sigma_{\rm total}$ = 3.7 mas/yr).  The dispersions in 
these cases actually measure a combination of each population's intrinsic velocity dispersion 
as well as scatter produced 
from instrumental errors.  The latter is clearly evident as we get smaller dispersions for 
brighter magnitude cuts which eliminate the lowest signal-to-noise ratio objects (the above 
numbers however do reflect the dispersions measured from using the entire data set).  A 
detailed analysis of the intrinsic dispersion, after correcting for instrumental scatter, 
will be provided in Saviane et al.\ (2003).  The extragalactic objects in our data, which 
have not moved during the 6 year baseline, are also convolved within the field clump 
identified above.  These are identified with red dots in the top-right panel.  

The CMD (left panel) shows the remarkably tight M4 main-sequence extending 
down to at least $V \sim$ 27 on top of the foreground/background stars.  A rich 
cluster white-dwarf cooling sequence is also seen stretching from 23 $\leq V \leq$ 
29.5.  As mentioned earlier, the line of sight through M4 ($d_{\rm M4}$ = 1.7 kpc) 
also intersects the spheroid of our Galaxy at a projected distance of 
2.2 kpc above the Galactic nucleus (for a tangent point distance of 7.6 kpc).  This 
population is more easily seen when the M4 main-sequence stars are removed by 
proper-motion selection (right panel).   The field population clearly extends to 
$V \sim$ 30 and shows some evidence for potential white dwarfs in the faint-blue region 
of the CMD.  These can be clearly confused with the galaxies in the $V$, $V{\rm-}I$ 
plane, thereby illustrating the importance of stellarity measurements.

The main sequence of the field population in the range $V$ = 24--30 is 
too broad to be consistent with the color distribution expected from a population with 
no metallicity spread, at the distance of the spheroid.  However, we can rule out a 
contribution of stars in these data from the Galactic 
bulge given that the line of sight passing through M4 lies well outside the infrared 
bulge as imaged by {\sl COBE} \rm \cite{arendt}.  The field also misses the 
highest surface-brightness isophotes in Binney, Gerhard \& Spergel (1997) and clearly 
avoids the classical metal-rich bulge population (McWilliam \& Rich 1994; Frogel, 
Tiede \& Kuchinski 1999).  Similarly, the redder stars in the field cannot be stars in 
the tri-axial bar near the Galactic center.  The orientation of the bar is such 
that the near side of the semi-major axis ($\sim$2 kpc) lies in the first quadrant.  
The angle between the Sun-center line and the bar's major axis in the plane of the 
disk is $\phi \sim$ 20$^{\rm o}$ \cite{gerhard}.  The line of sight through M4 
to the center of the Galaxy, ($l$, $b$) = (350.97$^{\rm o}$, 15.97$^{\rm o}$) also 
falls in this quadrant.   Given the thickness of the bar and axis ratio 1.0:0.6:0.4 
(Binney, Gerhard \& Spergel 1997), an azimuthal angle of 20$^{\rm o}$ would place the 
near side of the bar major-axis only $\sim$6.5 kpc from us (assuming an 8 kpc 
Galactocentric distance).  However, with a latitude of $\sim$16$^{\rm o}$, the 
line of sight through M4 is already 1.9 kpc above the plane at a distance of 6.5 
kpc and therefore well above the thickness of the bar.  Other bar models, such as 
that in Cole \& Weinberg (2002), would also not intersect our line of sight.  

A better explanation for the thickness of the main sequence is that a small 
admixture of metal-rich thick disk stars along the line of sight make up 
the redder population.  This is expected in these data given that 
the scale height of the thick disk is 1-1.5 kpc (Kuijken \& Gilmore 1991; 
Bienaym\'{e}, Robin \& Cr\'{e}z\'{e} 1987).  These would be co-rotating, 
perhaps with a small lag, and therefore reside closer to the 
location of the galaxies. However, the proper-motion diagram does not show any 
obvious evidence of this population even when isolating only the reddest, brightest 
possible thick disk stars.  The exact interpretation of the thickness of the 
observed main sequence is therefore still somewhat unclear and currently under 
investigation.

\section{White Dwarfs} \label{analysis}

\subsection{The Reduced Proper Motion Diagram} \label{rpmd}

With galaxies clearly separated from stars in these data, finding spheroid/dark 
halo white dwarfs should be relatively easy considering they must be faint, blue and 
have a proper motion consistent with the spheroid clump.  Since we also may have 
contamination from the thin and thick disk along our line of sight, the best way to 
isolate the inner-halo white dwarfs is to select them from the reduced proper motion 
diagram (RPMD).  This two dimensional plot ($H_{V}$ vs $V{\rm-}I$, where $H_{V}$ = $V$ + 
5log($\mu_{\rm total}$) + 5, and $\mu_{\rm total}$ is measured in $''$/yr) removes the 
different distances of individual 
populations by using the size of the proper motion as an effective distance indicator 
to offset the apparent magnitude.  Therefore, similar populations, despite their 
distance, will occupy similar regions on this diagram based both on their stellar 
properties and kinematics.  Figure 5 presents our stellar objects (stellarity $>$ 
0.7) for the large field clump (see Figure 4) on this plane together with the 
M4 stars (red dots).  

The RPMD clearly shows that the majority of our spheroid main-sequence sample (small 
dark dots) is similar to M4 (red dots).  This is expected considering the mean 
properties, such as the age and metallicity, are similar for both M4 and the spheroid.  
For a fixed magnitude, we see objects both above and below the M4 main-sequence stars 
on this plane.  The objects above represent those with smaller proper motions (Pop I) 
and those below have greater proper motions (Pop II).  The M4 white-dwarf cooling 
sequence is seen towards the blue end of the diagram and represents a locus for potential 
halo white dwarfs.  Several candidate white dwarfs from the field population (larger 
dark dots) are clearly seen sprinkled through the range 14 $\lesssim H_{V} \lesssim$ 25, 
near $V{\rm-}I \sim$ 1.  Most of these objects are found above the M4 white dwarfs, 
suggesting they have smaller proper motions (disk objects).  Three objects are found 
below the M4 white dwarfs suggesting possible spheroid stars.  Also shown are the locations 
of the thick-disk and putative dark halo-white dwarfs from the Mendez (2002) study 
(green squares) and those from the Nelson et al.\ (2002) study (blue triangles).  The 
faintest Mendez object as well as the faintest two Nelson objects agree very well in 
location with the M4 white dwarfs on the RPMD (these are the claimed dark halo detections, 
see \S \ref{discussion} for further discussion).  The remaining Mendez and Nelson objects 
are found above the M4 stars, suggesting a possible thick disk origin.  These points 
have all been corrected to match the reddening and extinction along this line of sight.

We also note that we do not select white dwarfs based on their location in the 
color-magnitude diagram for a reason.  One would naively expect that the spheroid 
white dwarfs should be consistent with the location of the M4 white dwarf cooling 
sequence shifted down to the tangent point (3.2 magnitudes fainter).  However we need 
to consider that the spheroid white dwarfs will occupy some depth in distance around the 
tangent point so there are both closer and farther objects.  Since the luminosity function 
\cite{hansen} of white dwarfs at the tangent point rises sharply at a point fainter than 
our limiting magnitude, we can expect to be systematically biased to seeing 
some fraction of the fainter white dwarfs within the region of the rising 
luminosity function slope if they reside closer to us than the tangent point.  
Since these fainter white dwarfs are redder than their brighter counterparts, 
the locus of these points will shift the +3.2-magnitude fiducial up to brighter 
magnitudes, on a parallel slope to the white dwarf cooling sequence.  Therefore, we 
can not use a color cut on the CMD to select these objects.  The approach used below 
does not suffer from these effects.

\subsection{Simulations} \label{simulations}

In order to assign our candidate white dwarfs to particular Galactic components, we need 
to model the underlying white dwarf populations. We model four such components; 
the thin disk (a double exponential with radial scale length 3~kpc and vertical scale 
height 0.3~kpc), thick disk (a double exponential with radial scale length 3~kpc and vertical scale 
height 1~kpc), spheroid/stellar halo (a power law with $\rho \propto r^{-3.5}$, where $r$ is the 
spherical Galactocentric radius) and a putative `dark halo' (a power law with $\rho \propto r^{-2}$).  
The Solar-neighborhood (white dwarf only) normalizations for the various components are taken to be 
$3.4 \times 10^{-3} M_{\odot}/pc^{3}$ (Holberg, Oswalt \& Sion 2002) for the thin disk, 
$10^{-4} M_{\odot}/pc^{3}$ for the thick disk (Oppenheimer et al.\ 2001; Reid, Sahu \& Hawley 2001), 
$3 \times 10^{-5} M_{\odot}/pc^{3}$ for the spheroid (Gould, Flynn \& Bahcall 1995) and $1.4 
\times 10^{-3} M_{\odot}/pc^{3}$ for the dark halo (corresponding to a fraction 20\% of the 
dark matter density, as indicated by the MACHO group's results, Alcock et al.\ 2000). 

The above density laws are projected onto the line of sight, taking into account the 
increasing volume at larger distance (for fixed solid angle). Using the same white dwarf 
models as in Hansen et al.\ (2002), we derive Monte-Carlo realizations of the field white 
dwarf populations distributed along our line of sight. The thin, thick and spheroid populations 
are drawn from a population with progenitor IMF $dN/dM \propto M^{-2.6}$. The dark halo 
population is drawn instead from a population with a Chabrier IMF (Chabrier 1999; IMF2 in the 
notation of that paper). The models of Hurley, Pols \& Tout (2000) were used to provide 
main-sequence lifetimes. The thin-disk population was assumed to be forming stars at a continuous 
rate (over the past 10~Gyr) while all the others were assumed to be 12~Gyr old bursts.  
The Monte-Carlo calculations include the detection probability using the incompleteness 
corrections from Richer et al.\ (2002) and Hansen et al.\ (2002).

The final results indicate that, on average, we expect 7.9 thin-disk, 6.3 thick-disk 
and 2.2 spheroid white dwarfs. The dark halo will contribute 2.5 white dwarfs if 20\% of 
the dark halo is made up of DA white dwarfs. The first three (i.e., the standard) populations 
are dominated by bright white dwarfs located near the tangent point (even for the disk 
populations, as the radial exponential largely compensates for the decrease due to the 
vertical exponential, at least for this line of sight). The dark-halo population, on 
the other hand, is dominated by objects at distances $\sim$2-3 kpc, a consequence of 
the bias inherent in the Chabrier IMF towards higher progenitor masses and thus 
shorter main-sequence lifetimes (corresponding to longer white dwarf incarnations). This 
results in an anticipated systematic color difference between any dark-halo white dwarfs 
(relatively red) and those from standard populations (relatively blue). We stress this 
applies particularly to our data set and does not necessarily extend to wider-field 
surveys.

\subsection{Results} \label{WD}

In Figure 6, we display the location of the white dwarfs in the RPMD.  We also 
overlay the positions of the simulated sample of white dwarfs (open circles) on this 
diagram (for a much larger area than that observed). The thin-disk 
(top-left) and thick-disk (top-right) stars are generally found brighter than the 
spheroid (bottom-left) stars, with a limit at $H_{V} \sim$ 25 beyond which there are 
very few disk white dwarfs.  A similar threshold 
was used by Flynn et al.\ (2001) to separate disk and halo white dwarfs.  They found 
that most dark halo white dwarfs would lie at 24.3 $\leq H_{V} \leq$ 26.3 and peak at 
$H_{V} \sim$ 25.3 (corrected to the extinction along this line of sight).  The {\em 
relative} numbers of simulated thin-disk, thick-disk and spheroid objects 
plotted in Figure 6 have been scaled to match the expectations in our data.  The $H_{V}$ 
$\sim$ 14 cutoff at the top of the diagram indicates the saturation limit of our data.  The 
simulated dark-halo white dwarfs (open circles, bottom-right) are clearly distinct 
from these conventional populations and are found at redder colors (for an explanation 
see \S \ref{simulations}).  Also included on this panel are those objects which did not 
satisfy the stellarity or proper motion cut (small crosses) to show that the data do 
extend fainter than $H_{V}$ = 25 and well into the dark halo regime.  

Although it is difficult to separate thin-/thick-disk white dwarfs, we can conclude 
that nine of our objects are consistent with disk white dwarfs ($H_{V} \leq$ 23, 
$V{\rm-}I \leq$ 1.5).  Three objects in our data ($H_{V}$ = 23.15, $V{\rm-}I$ = 0.48, 
$H_{V}$ = 23.78, $V{\rm-}I$ = 0.93, and $H_{V}$ = 24.09, $V{\rm-}I$ = 0.93) which are 
fainter than this cut are found within the tail of the thick disk simulated sample.  These 
objects are however in excellent agreement with the simulated spheroid white dwarfs 
(bottom-left).  These are also the three objects which fall below the M4 white dwarfs in 
Figure 5.  Although the predicted number 
of stars allows for incompleteness in the data set, we must still correct these 
observed numbers of stars for the incompleteness caused by mismatches in the SExtractor vs 
ALLSTAR classifications.  In \S \ref{expectedgalaxies} we estimated 25\% of the ALLSTAR 
sources were missed for these magnitude bins.  Our final corrected numbers of detected 
objects are therefore 12 disk white dwarfs (14 expected) and 4 spheroid white dwarfs 
(2 expected) and therefore in good agreement with the predictions considering that the 
observed spheroid number is an upper limit (i.e., two of the objects are consistent 
with the tail of the thick disk distribution).

A key result from Figure 6 is that the simulated dark halo region is basically not populated 
with data points.  Only one object which we classified as thick disk, $H_{V}$ = 22.25, 
$V{\rm-}I$ = 1.38, marginally agrees with the edge of the simulated sample (small open circles).  
Less than 4\% of our simulated dark-halo sample is 
found with a reduced proper motion less than this object, suggesting $<$0.1 expected 
real objects.  Also, the color of this star falls close to the blue edge of the simulated 
sample and is only marginally consistent with the expected location of dark-halo white dwarfs.  
A possible explanation for the other extremely red object, as well as possibly this object, 
is that these may be binary thin-/thick-disk white dwarfs with red main-sequence companions.  
We can also rule out the spheroid candidate objects as potential dark-halo white dwarfs given 
that they have much bluer colors.  We've added photometric and astrometric error bars to all 
candidates in the bottom-right panel of Figure 6.  The errors are 1-$\sigma$ values from the 
results of artificial star tests.  For example, the magnitude and color errors are the input 
vs output photometric values and the positional errors are the input vs output coordinates 
in the tests.  The $H_{V}$ error combines the contribution from the magnitude and 
proper-motion terms.  We have left these error bars off from other panels to prevent 
cluttering in the diagrams. We also note that just relaxing the stellarity cut to 0.5 
would not allow any of the spurious sources (small crosses) to qualify as candidate dark 
halo white dwarfs given their inconsistent proper motions.  Finally, we indicate the limit 
at $H_{V} \sim$ 27 above which we would have expected dark-halo white dwarfs given our 
simulations and detection efficiency.  Table 2 presents our final white dwarf candidate 
list, both for the disk and spheroid.

\section{Discussion}\label{discussion}

Contrary to our result, recent claims by Oppenheimer et al.\ (2001), Koopmans \& Blandford 
(2001), Nelson et al.\ (2002) and Mendez (2002) suggest that a halo white-dwarf population 
exists and that the number of white dwarfs is in excess of that expected from the stellar 
spheroid alone.  Others (Reid, Sahu \& Hawley 2001; Hansen 2001; Richer 2001) have argued 
that this small population is a component of the thick disk.  Gibson \& Mould (1997) used 
simple chemical evolutionary arguments to show that it is difficult to reconcile the 
observed spheroid main-sequence chemical abundance with that expected from an IMF which would 
give the speculated halo white-dwarf density.  We briefly summarize these independent studies 
and place them in the context of the present observations.  

Oppenheimer et al.\ (2001) sampled 4165 sq.\ deg.\ towards the Southern Galactic Cap and found 
38 cool white dwarfs, the majority of which they assign to a dark-halo population.  
The data from this study were made available and consequently several authors have 
re-analyzed the result.  Koopmans \& Blandford (2001) used a dynamical model and performed 
a maximum-likelihood analysis on the Oppenheimer sample of white dwarfs and found 
two kinematically distinct populations at a greater than 99\% confidence limit 
(a thick disk and a flattened halo).  However, Reid, Sahu \& Hawley (2001) compared 
the kinematics of the Oppenheimer sample to the local M-dwarf sample and 
suggested that the majority of this population resides in the thick disk.  
Richer (2001) and Hansen (2001) examined the temperature and age distribution of 
the Oppenheimer white dwarf sample and found that the star-formation history for 
most of these stars is more representative of the disk rather than the halo.  

The Nelson et al.\ (2002) study looked for high-proper-motion objects in the 
Groth-Westphal strip.  They measured proper motions over a 7-year baseline 
using WFPC2 observations of a 74.8 arcmin$^{\rm 2}$ region.  Of the 24 high-proper-motion 
objects detected, five are believed to be strong white dwarf candidates 
(two of these are dark halo, three are disk) and a further two are classified as marginal 
candidates.  The two halo objects are the faintest two blue triangles in Figure 5 and the 
three disk objects are the brightest three blue triangles in Figure 5.  By creating a model 
of the stellar components of our Galaxy, Nelson et al.\ (2002) demonstrate that the 
observations are clearly in excess of the expected number of white dwarfs from 
the Galactic stellar components.  Although systematic errors due to uncertainties 
in the model parameters are large, the observations suggest a 7\% white-dwarf dark 
halo.  

Most recently, Mendez (2002) has presented new evidence for a bi-modal 
kinematic population of old white dwarfs and suggests that most of the 
required dark matter in the Solar vicinity can be accounted for by 
these populations.  This analysis is based on a proper-motion 
membership study of the Galactic globular cluster NGC\,6397 
\cite{king}.  Mendez (2002) used both the reduced proper motion 
and the colors of field objects to assign six white-dwarf 
candidates to the thick disk and one to the dark halo (the faintest 
green square shown in Figure 5).  The Mendez (2002) study 
gives no information on the morphology of the sources and possible 
extragalactic contamination.

There is clearly no convincing answer yet as to whether white dwarfs make up an important 
component of the dark matter in our Galaxy.  The controversy spurred from these 
studies can only be resolved by obtaining deeper, larger images of halo fields which 
could contain many white dwarfs.  The current data set represents an important 
constraint in the context of searching for these objects as it is sensitive 
to white dwarfs from all three (disk, spheroid and dark-halo) components and it is only 
the dark-halo ones that are not found.  Furthermore, our images represent the deepest-ever 
probe into the inner halo of our Galaxy and we do see the stellar halo main-sequence 
progenitors of the white dwarfs.  By using only directly observable quantities, such as 
the proper motions, magnitudes and colors we have demonstrated that these data do 
not support a white-dwarf component to the dark halo.

\section{Conclusions} \label{conclusions}

A 6-year baseline of imaging of the globular star cluster M4 has 
allowed us to separate out the cluster stars from the background spheroid 
population.  Over this time, distant extragalactic sources have 
not moved and are identified using image morphology criteria.  We establish a 
zero-motion frame of reference using these galaxies and search for halo white 
dwarfs based on their morphology and kinematics.  We find that distinguishing faint galaxies 
from field stars is impossible when based solely on proper motions, and that an index of
stellarity is crucial in separating the two classes.  Based on the reduced 
proper motion diagram, we identify nine thin/thick disk and three spheroid 
white dwarfs in these data.  These numbers are consistent with the expected 
contribution of the conventional populations along the line of sight.  
Additionally, 2.5 dark halo white dwarfs are expected in these data based on 
a 20\% white dwarf dark halo and are not found.  Therefore, we do not need to 
invoke a dark halo population of white dwarfs to explain the present data.

The extragalactic reference frame also allows us to perform two other 
important measurements.  First, this fixed frame gives us an independent 
measurement of the fundamental Galactic constant, $\Omega_\circ$ = 
$\Theta_\circ/R_\circ$ = 25.3 $\pm$ 2.6 km/s/kpc.   This provides a 
velocity of the Local Standard of Rest $v_{\rm LSR}$ = $\Theta_\circ$ = 202 $\pm$ 
25 km/s at $R_\circ$ = 8.0 $\pm$ 0.5 kpc, in agreement with independent studies.  
Secondly, the galaxies give us a direct measurement of M4's absolute proper 
motion, $\mu_{\alpha}$\alphab = $\rm-$12.26 $\pm$ 0.54 mas/yr, 
$\mu_{\delta}$\deltab = $\rm-$18.95 $\pm$ 0.54 mas/yr, also in good 
agreement with the latest studies.  

\begin{acknowledgements}
The author wishes to thank C.A. Nelson et al.\ for making available their simulated 
white dwarfs and D.I. Dinescu for useful information about coordinate conversions.  
We also acknowledge I. King \& S. Courteau for useful discussions 
regarding the circular speed curve and P. Guhathakurta for pointing out an important 
oversight in a preliminary version of this paper.  JSK received financial support 
during this work through an NSERC PGS-B graduate student research grant.   HBR is 
supported in part by the Natural Sciences and Engineering Research Council 
of Canada. HBR also extends his appreciation to the Killam Foundation and the Canada 
Council for the award of a Canada Council Killam Fellowship. RMR and IS acknowledges support 
from proposal GO-8679 and BMH from a Hubble Fellowship HF-01120.01 both of which were 
provided by NASA through a grant from the Space Telescope Science Institute which is 
operated by AURA under NASA contract NAS5-26555.  BKG acknowledges the support of the 
Australian Research Council through its Large Research Grant Program A00105171.
\end{acknowledgements}

\newpage

\figcaption[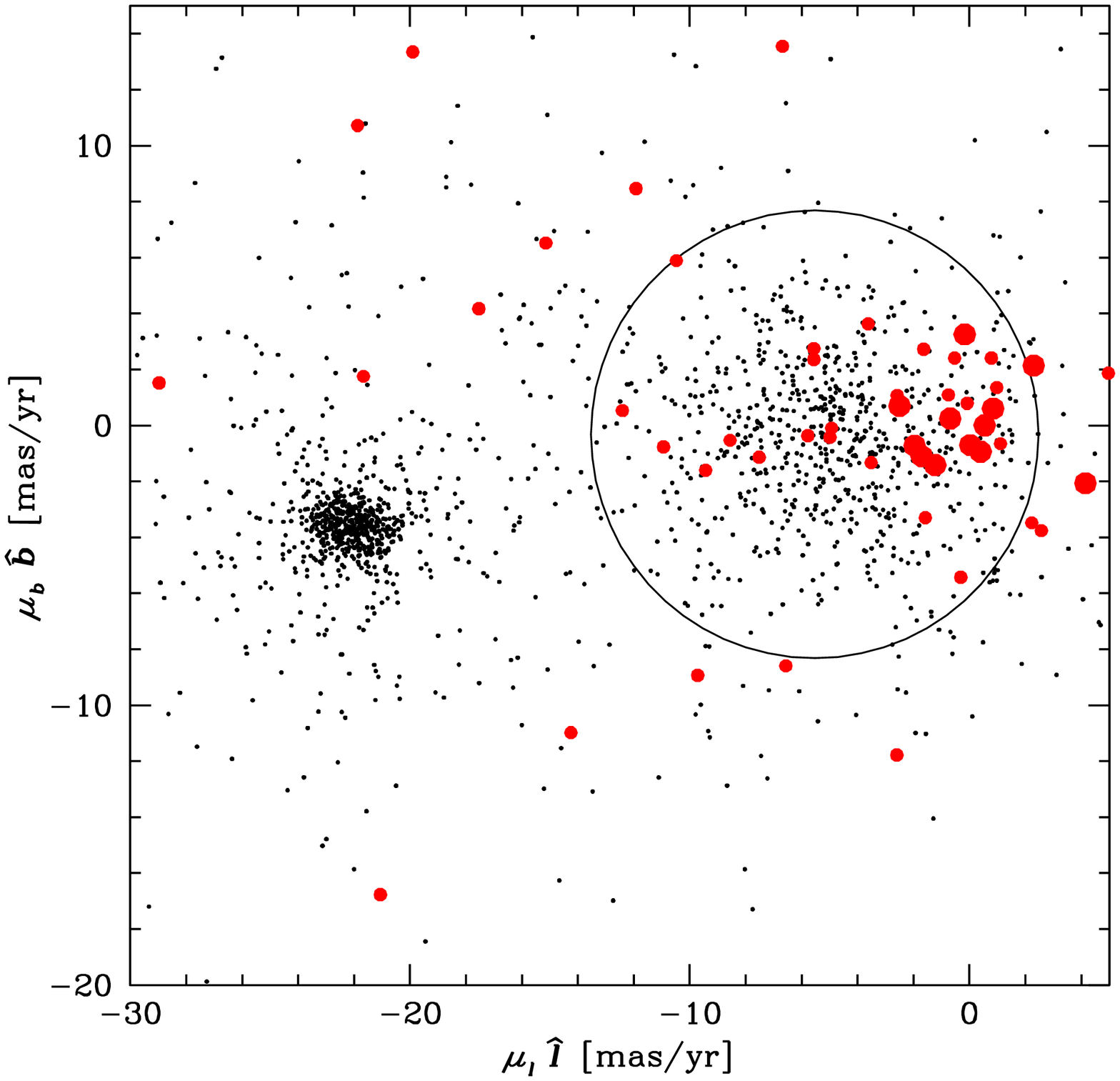]{Figure 1: The proper-motion diagram for all 
stars ($\mu_l{\bf \hat{l}}$, $\mu_b{\bf \hat{b}}$) is shown with galaxies represented 
as red dots (see \S \ref{galaxies}).  The vector proper motions are 
expressed using the convention where ${\bf \hat{l}}$ and ${\bf \hat{b}}$ are 
unit vectors in the corresponding directions.  The galaxies used to determine 
the zero-motion reference frame are displayed as larger dots (see \S 
\ref{centering}).  The tighter clump of M4 stars is clearly distinct 
from the more diffuse spheroid clump.}

\figcaption[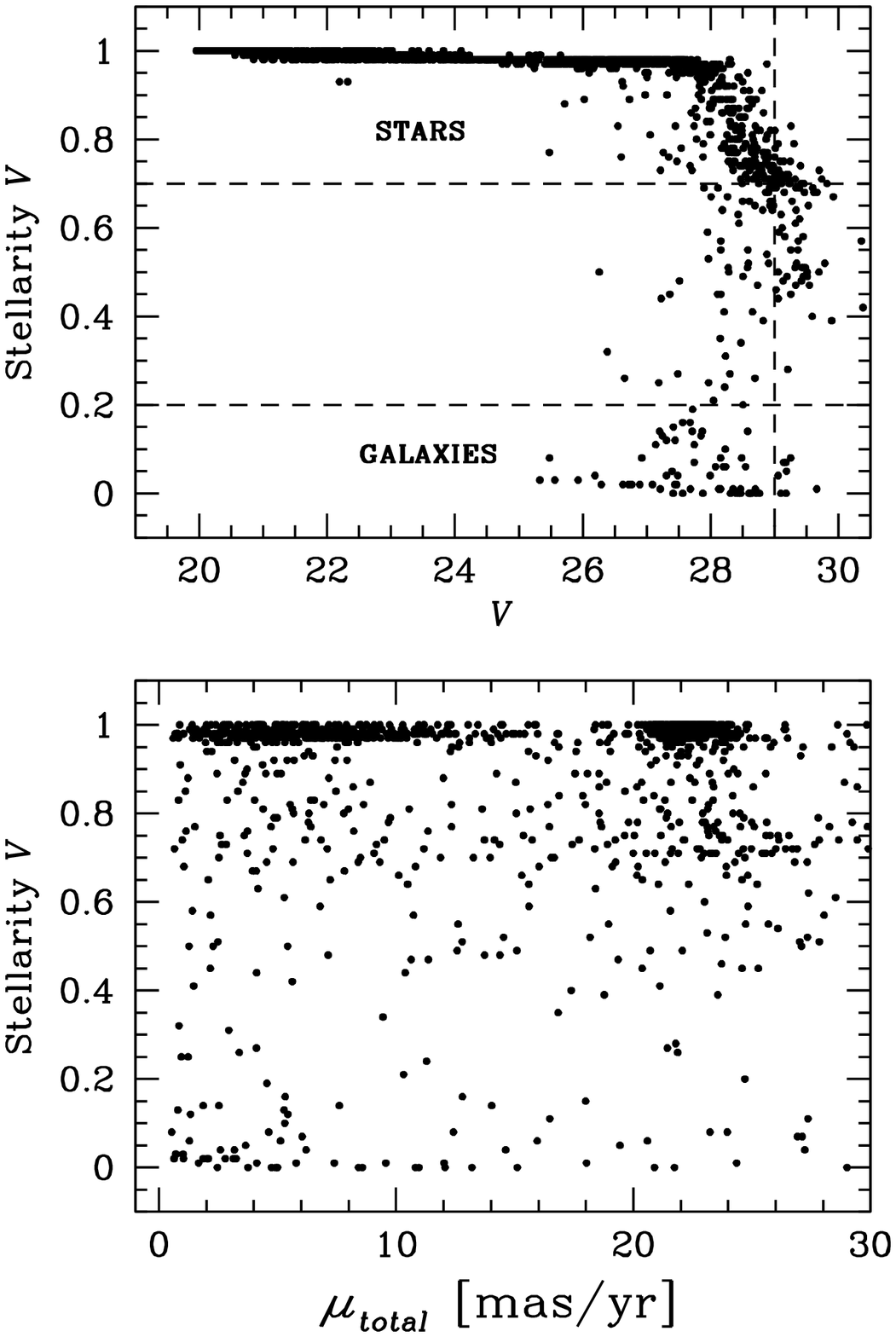]{Figure 2: {\it Top} \rm - A plot of stellarity vs magnitude 
shows that we have clearly separated stars from galaxies down to $V \sim$ 29.  
{\it Bottom} \rm - The stellarity of all objects is plotted against the total proper-motion 
displacement ($\mu_{\rm total}$ = $\sqrt{(\mu_{l}{\bf \hat{l}})^{2} 
+ (\mu_{b}{\bf \hat{b}})^{2}}$) to illustrate the different populations along the 
M4 line of sight (see \S \ref{expectedgalaxies}).}

\figcaption[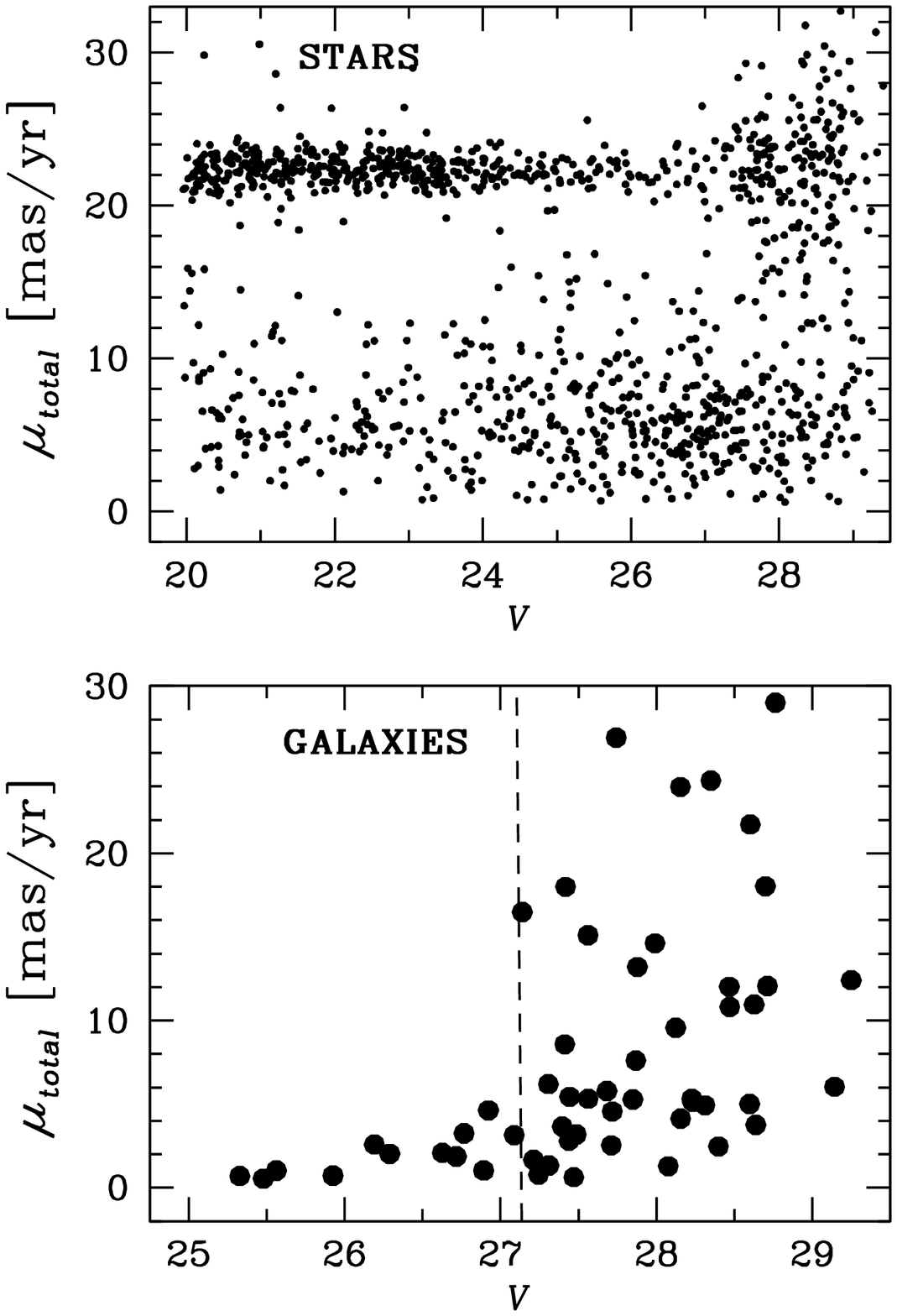]{Figure 3: The total proper motion displacement is plotted 
as a function of $V$ magnitude for both stars ({\it top}) and galaxies ({\it bottom}) 
as determined by SExtractor.  The stars are found along two sequences, representing 
the cluster ($\mu_{\rm total} \sim$ 22.5 mas/yr) and the spheroid ($\mu_{\rm total} 
\sim$ 5 mas/yr) populations.  The galaxies are confined to zero motion for $V \lesssim$ 27, 
beyond which their astrometry degrades (see \S \ref{centering}). These brightest galaxies 
are used to define the zero-motion frame of reference.}

\figcaption[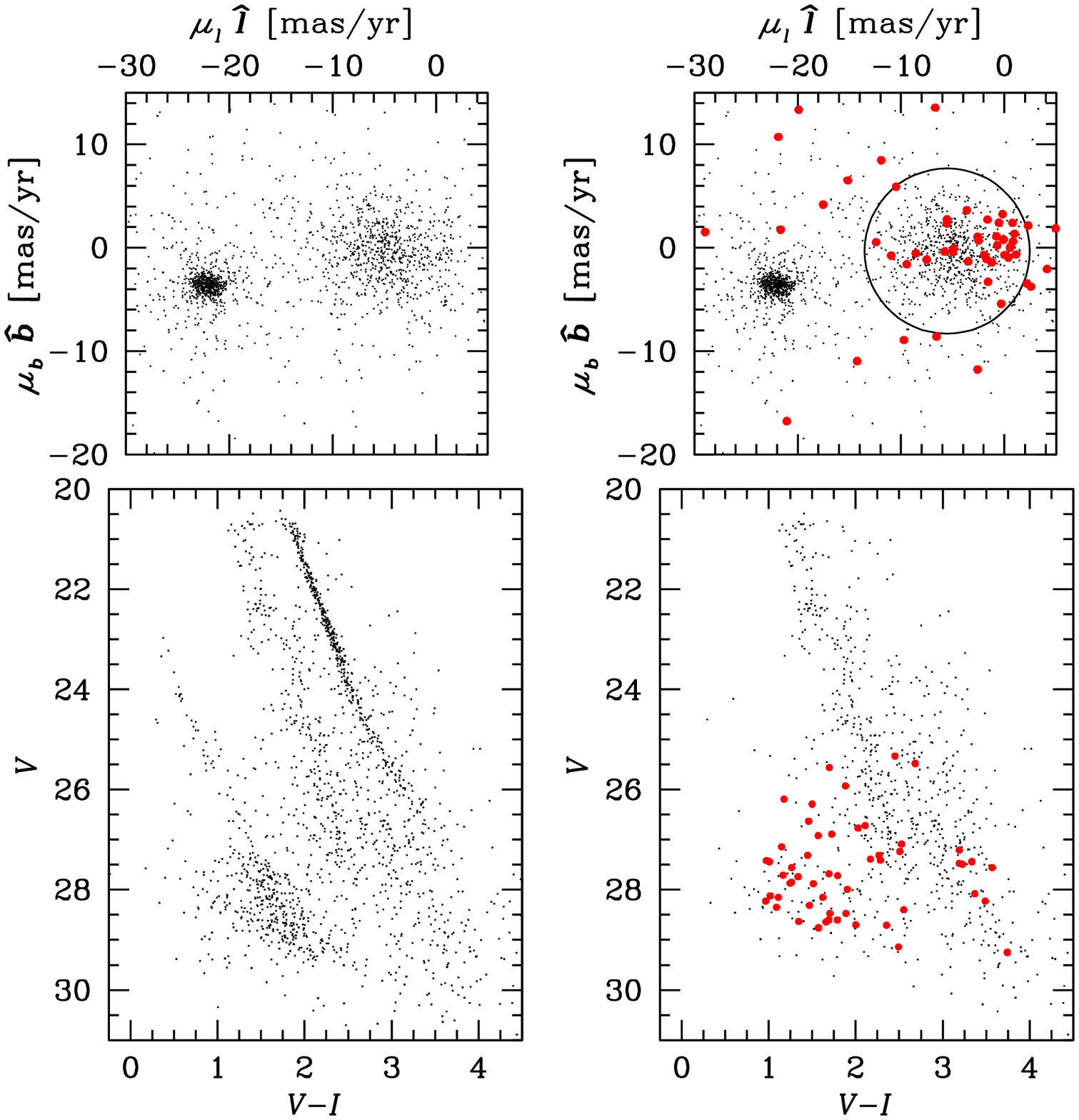]{Figure 4: {\it Top} \rm - The proper-motion diagram for all 
stars ($\mu_l{\bf \hat{l}}$, $\mu_b{\bf \hat{b}}$) is shown with galaxies represented as red dots (see 
\S \ref{galaxies}).  The tighter clump represents M4 members and the more diffuse 
clump predominantly represents the background spheroid stars. {\it Bottom} \rm - The 
CMD for all objects in the image is shown (left panel) and for only those objects which 
fall within the spheroid field clump (right panel).  The galaxies are shown in red.}

\figcaption[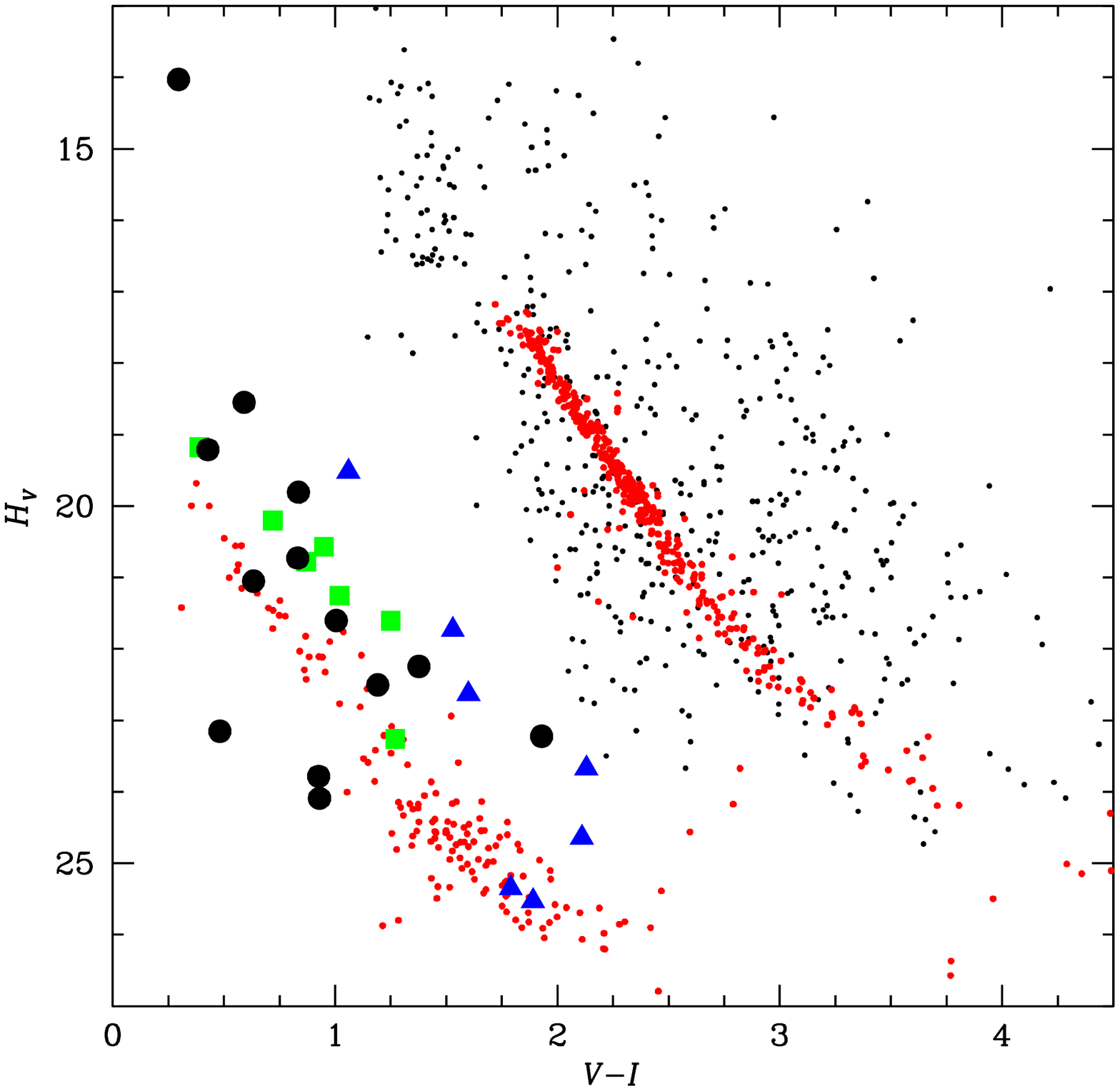]{Figure 5: The reduced proper motion 
diagram ($H_{V}$ = $V$ + 5log($\mu^{''}_{\rm total}$) + 5) is shown for all 
field stars with stellarity $>$ 0.7 and M4 members (red dots).  Inner halo white 
dwarf candidates are displayed with larger circles.  White dwarfs from the 
Mendez 2002 study (green squares) and the Nelson et al. 2002 study (blue triangles) 
are also displayed after correcting for reddening and extinction differences.}

\figcaption[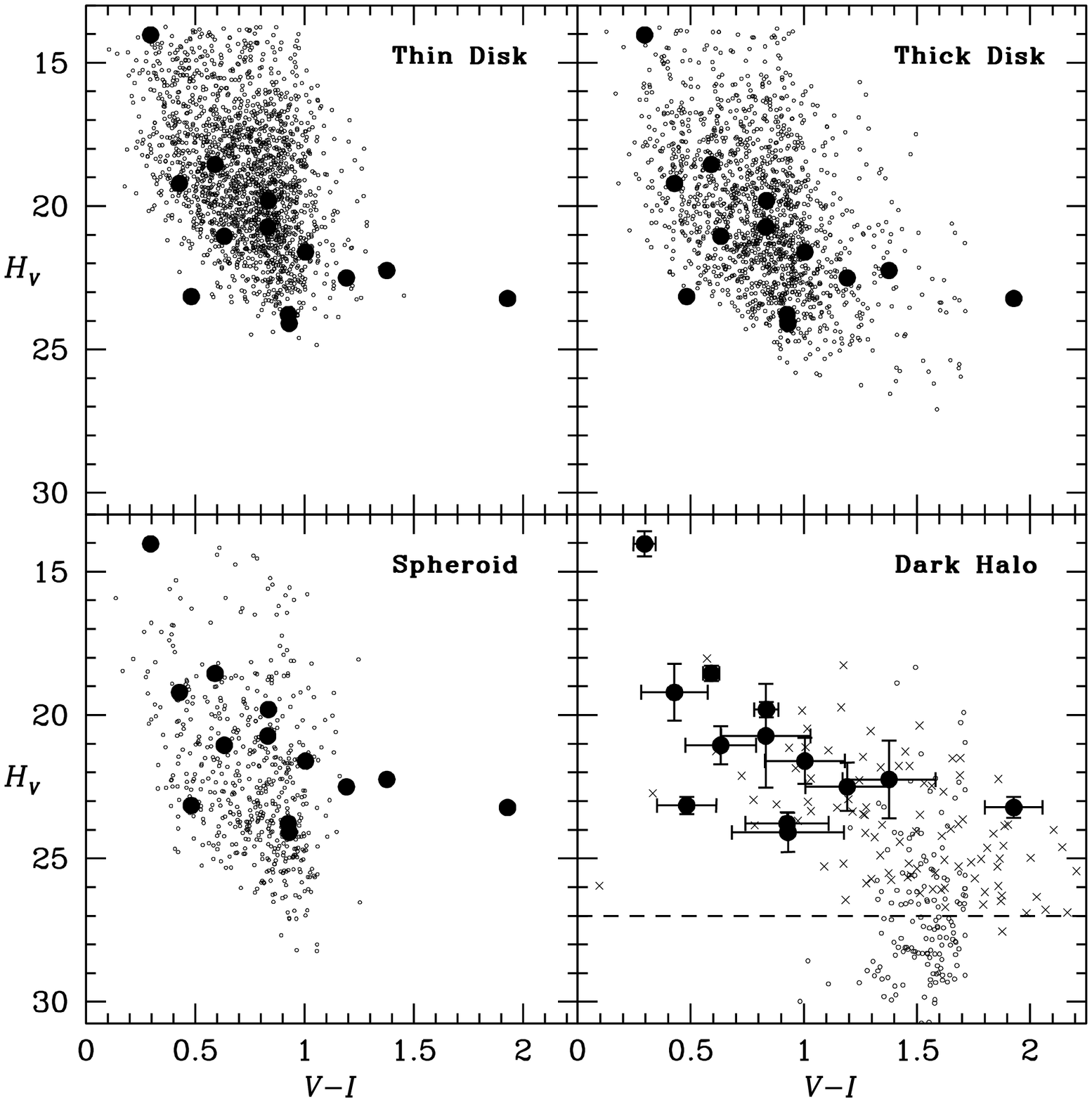]{Figure 6: A simulated sample of stars belonging 
to the thin disk (top-left), thick disk (top-right), spheroid (bottom-left) and dark halo 
(bottom-right) are shown on the RPMD as small open circles in all panels.  Candidate white dwarfs 
from the present data are shown as large filled circles.  After correcting for mismatches 
in the SExtractor vs ALLSTAR photometry, we find 12 (9 originally) thin/thick disk 
white dwarfs and 4 (3 originally) spheroid white dwarfs.  Although 2.5 
objects are expected from a 20\% dark halo dominated by white dwarfs, we find 
none in these data (see \S \ref{WD}).  1-$\sigma$ astrometric and photometric error bars 
for all candidates are only shown in the bottom-right panel to prevent cluttering in other 
panels.  Small crosses depict those objects which did not satisfy the stellarity or 
proper motion cut and show that the data set is sensitive to $H_{V} \lesssim$ 27 (and 
therefore to dark halo white dwarfs).}

\newpage
\epsscale{1.0}
\plotone{f1.eps}

\newpage
\epsscale{1.0}
\plotone{f2.eps}

\newpage
\epsscale{1.0}
\plotone{f3.eps}

\newpage
\epsscale{1.0}
\plotone{f4.eps}

\newpage
\epsscale{1.0}
\plotone{f5.eps}

\newpage
\epsscale{1.0}
\plotone{f6.eps}

\newpage

\begin{deluxetable}{lll}
\tabletypesize{\scriptsize} \tablecaption{Cluster, Field Parameters and Observational 
Data \label{table1}} \tablewidth{0pt}
\tablehead{} \startdata
Field Location: & \\
$\alpha_{\rm J2000}$  & RA                   & = 16$^{\rm h}$23$^{\rm m}$54.6$^{\rm s}$  \\ 
$\delta_{\rm J2000}$ & declination          & = $\rm-$26$^{\rm o}$32$'$24.3$''$          \\
l$_{\rm J2000}$       & Galactic longitude   & = 351.01$^{\rm o}$              \\   
b$_{\rm J2000}$       & Galactic latitude    & = 15.91$^{\rm o}$               \\ 
\\
Cluster Distance \& Reddening\tablenotemark{1} : & \\
$(m {\rm-}M)_{V}$   & apparent distance modulus   & = 12.51     \\
$E(B {\rm-}V)$      & reddening                   & = 0.35      \\
$A_{V}$       & visual extinction               & = 1.33      \\
$(m {\rm-}M)_\circ$   & true distance modulus       & = 11.18      \\
$d$           & distance from Sun         & = 1.72 kpc  \\
\\
Background Field (Spheroid) Distance \& Reddening: & \\
$(m {\rm-}M)_{V}$   & apparent distance modulus   & = 15.71     \\
$E(B {\rm-}V)$      & reddening (all in front of M4)  & = 0.35      \\
$A_{V}$       & visual extinction               & = 1.33      \\
$(m {\rm-}M)_\circ$   & true distance modulus       & = 14.39      \\
$d$           & tangent point distance          & = 7.6 kpc  \\
$z$           & projected distance from Nucleus & = 2.2 kpc \\
\\
Metallicity (Cluster\tablenotemark{2} \ \& Spheroid: & \\
$[Fe/H]$             & heavy metal abundance         & = $\rm-$1.3      \\
\\
\tableline
Data (GO 5461\tablenotemark{3} \ - cycle 4 \& GO 8679 - cycle 9): & \\
No. of Images -- F555W - 1995      & 15$\times$2600 seconds   \\
No. of Images -- F606W - 2001      & 98$\times$1300 seconds \\
No. of Images -- F814W - 1995      & 9$\times$800 seconds \\
No. of Images -- F814W - 2001      & 148$\times$1300 seconds \\
\\
Limiting Magnitude (Based on SExtractor Classifications): & \\
$V$             & 29      \\
$I$             & 27.5    \\
\\

\enddata
\tablerefs{1. Richer et al (1997), 2. Djorgovski (1993), 3. Ibata et al (1999)} 

\end{deluxetable}

\newpage

\begin{deluxetable}{lcccccccc}
\tabletypesize{\scriptsize} \tablecaption{Disk and Spheroid White Dwarfs 
\label{table2}} \tablewidth{0pt}
\tablehead{\colhead{Candidate} & \colhead{Designation} & 
\colhead{$\alpha_{\rm J2000}$} & \colhead{$\delta_{\rm J2000}$} & \colhead{$V$} & 
\colhead{$V{\rm -}I$} & \colhead{$\mu$ (mas/yr)\tablenotemark{1}} & \colhead{$\mu_l$${\bf \hat{l}}$ 
(mas/yr)} & \colhead{$\mu_b$${\bf \hat{b}}$ (mas/yr)}} \startdata

WD 1        & disk     & 16:23:52.76 & $\rm-$26:33:13.07 &  24.60  &  0.30 & 0.77    &       0.77   & $\rm-$0.05  \\
WD 2        & disk     & 16:23:54.49 & $\rm-$26:31:19.38 &  27.17  &  0.43 & 2.56    &       2.33   &       1.06  \\
WD 3        & disk     & 16:23:55.54 & $\rm-$26:32:36.64 &  24.18  &  0.59 & 7.47    & $\rm-$7.27   & $\rm-$1.71  \\
WD 4        & disk     & 16:23:55.87 & $\rm-$26:33:03.28 &  27.43  &  0.63 & 5.31    & $\rm-$5.22   &       0.96  \\
WD 5        & disk     & 16:23:58.75 & $\rm-$26:32:40.13 &  25.22  &  0.84 & 8.28    & $\rm-$8.26   &       0.51  \\
WD 6        & disk     & 16:24:02.33 & $\rm-$26:32:25.55 &  27.98  &  0.83 & 3.54    &       0.88   &       3.43  \\
WD 7        & disk     & 16:24:02.81 & $\rm-$26:32:42.61 &  27.82  &  1.01 & 5.70    & $\rm-$5.37   & $\rm-$1.91  \\
WD 8        & disk     & 16:24:03.10 & $\rm-$26:32:45.10 &  27.99  &  1.19 & 7.99    & $\rm-$7.41   & $\rm-$3.01  \\
WD 9        & spheroid & 16:23:55.28 & $\rm-$26:32:47.47 &  27.02  &  0.48 & 16.85   & $\rm-$15.48  &       6.66  \\
WD 10       & spheroid & 16:23:55.87 & $\rm-$26:32:37.77 &  28.34  &  0.93 & 14.15   & $\rm-$13.69  &       3.57  \\
WD 11       & spheroid & 16:23:58.12 & $\rm-$26:32:10.13 &  27.82  &  0.93 & 15.58   & $\rm-$15.31  &       2.89  \\
WD 12       & disk?    & 16:23:53.25 & $\rm-$26:33:10.66 &  28.25  &  1.38 & 6.30    & $\rm-$4.73   & $\rm-$4.17  \\
WD 13       & disk?    & 16:23:59.35 & $\rm-$26:31:28.52 &  27.49  &  1.93 & 13.99   & $\rm-$13.89  & $\rm-$1.67  \\

\enddata

\tablenotetext{1}{$\mu_{\rm total} = \sqrt{(\mu_{l}{\bf \hat{l}})^2 + (\mu_{b}{\bf \hat{b}})^2}$} 



\end{deluxetable}

\end{document}